\documentclass[10pt]{iopart}
\usepackage{iopams}
\usepackage{graphicx}
\usepackage{cite}
\usepackage{stfloats}
\usepackage{subcaption}
\begin{document}

\title{Shape optimization of superconducting transmon qubit for low surface dielectric loss}

\author{Sungjun Eun, Seong Hyeon Park, Kyungsik Seo, Kibum Choi and Seungyong Hahn}

\address{Department of Electrical and Computer Engineering, Seoul National University, 1 Gwanak-ro,	Gwanak-gu, Seoul 08826, Republic of Korea}
\ead{pajoheji0909@snu.ac.kr and hahnsy@snu.ac.kr}
\vspace{10pt}
\begin{indented}
\item[]November 2022
\end{indented}

\begin{abstract}
	Surface dielectric loss of superconducting transmon qubit is believed as one of the dominant sources of decoherence. Reducing surface dielectric loss of superconducting qubit is known to be a great challenge for achieving high quality factor and a long relaxation time ($T_{1}$). Changing the geometry of capacitor pads and junction wire of transmon qubit makes it possible to engineer the surface dielectric loss. In this paper, we present the shape optimization approach for reducing Surface dielectric loss in transmon qubit. The capacitor pad and junction wire of the transmon qubit are shaped as spline curves and optimized through the combination of the finite-element method and global optimization algorithm. Then, we compared the surface participation ratio, which represents the portion of electric energy stored in each dielectric layer and proportional to two-level system (TLS) loss, of optimized structure and existing geometries to show the effectiveness of our approach. The result suggests that the participation ratio of capacitor pad, and junction wire can be reduced by 16\% and 26\% compared to previous designs through shape optimization, while overall footprint and anharmonicity maintain acceptable value. As a result, the TLS-limited quality factor and corresponding $T_{1}$ were increased by approximately 21.6\%.
\end{abstract}

\vspace{2pc}
\noindent{\it Keywords}: Shape optimization, superconducting qubits, surface dielectric loss, transmon.


\maketitle

\ioptwocol

\section{Introduction}
\indent Over the past decade, the dramatic improvement of the coherence time of superconducting qubits \cite{Devoret_superconducting} made the superconducting qubit a promising platform for quantum information processing, and recent demonstration of quantum supremacy using superconducting qubit \cite{Arute_quantum} made it even more attractive. Among various types of superconducting qubits, transmon qubit \cite{Koch_charge} is one of the most popular structures, which is a charge qubit featuring an exponential decrease of charge noise due to shunt capacitance connected to Josephson Junction (JJ). One of the biggest interests of quantum engineers is the coherence of superconducting qubit \cite{Krantz_quantum}, which is necessary to realize practical superconducting quantum computers. Achieving longer relaxation time ($T_{1}$) and dephasing time ($T_{2}$) is a matter of utmost importance. It is commonly believed that two-level system (TLS) defects \cite{Anderson_anomalous} contained in interface layers act as a noise source and limit quality factors in superconducting circuits \cite{Martinis_decoherence, Gao_semiempirical,Gao_experimental,Muller_towards, Geerlings_improving}. Since TLS defects are coupled to a qubit system via an oscillating electric field in quantum circuits \cite{Lisenfeld_electric}, TLS loss of transmon qubits can be calculated from the electric field profile and geometry of the superconducting qubit \cite{Wenner_surface,Wang_surface,Melville_comparison,Gambetta_investigating,Calcusine_analysis,Martinis_surface}. \\
\indent One of the common approaches to reducing TLS loss is changing material properties or fabrication methods since TLS loss is heavily dependent on material composition \cite{Melville_comparison,Wang_towards,Place_new, Deng_titanium}. Geometry dependence of TLS loss in a superconducting circuit was also studied in \cite{Geerlings_improving, Dial_bulk, Gambetta_investigating,Park_design}, and the potential trade-off between footprint and quality factor \cite{Deng_titanium} is commonly observed in those studies. However, to our best knowledge, shape optimization solely focusing on reducing participation ratio and TLS loss of transmon qubit was not vastly been studied yet. Despite the inherent uncertainty of TLS-related effects \cite{Melville_comparison,Lisenfeld_electric,Linsenfeld_decoherence}, we can reduce TLS loss of transmon qubit with shape optimization through proper setting of loss tangent and participation ratio calculation.\\
\indent Shape and topology optimization \cite{Park_design_book} has been studied profoundly in mechanical engineering \cite{Qian_topology,Allaire_structural,Deng_self}, antenna optimization \cite{Koziel_fast,John_antenna,Zhou_level,Lizzi_optimization,Lizzi_PSO,Hassan_topology,Toivanen_gradient,Koulouridis_novel}, and electromagnetic systems \cite{Park_design_book,Lee_continuum}. In particular, for planar antenna, it has been shown that essential aspects of antenna, such as impedance matching \cite{Hassan_topology}, bandwidth \cite{John_antenna,Lizzi_optimization,Lizzi_PSO}, and current density of dipole antenna \cite{Zhou_level}, can be optimized through shape optimization. Optimization techniques used in those studies include adjoint-based sensitivity analysis \cite{Hassan_topology,Zhou_level,Georgieva_feasible}, genetic algorithm (GA) \cite{Weile_genetic,Koulouridis_novel}, particle swarm optimization (PSO) \cite{Lizzi_optimization,Lizzi_PSO} and artificial neural network \cite{Misilmani_review}.\\
\indent In this paper, we present the shape optimization-based approach to reduce surface dielectric loss and participation of transmon qubit. Inspired by shape optimization techniques of planar antenna, we find a geometry that can achieve lower participation while the overall footprint is limited. Similar to spline-based planar antenna \cite{John_antenna,Lizzi_optimization,Lizzi_PSO,Koulouridis_novel}, we used a spline shape to express the geometry of the capacitor pad and junction wire, and a global optimization tool was utilized to find the optimal geometry. In \sref{Section2}, the fundamental scheme for participation ratio calculation and optimization is introduced. The effect of superconductor's surface impedance is also discussed to justify our simulation settings. In \sref{Section3}, the results of optimization including convergence, optimized geometry, and extracted key parameters are presented. At last, we give conclusions on this study by discussing the increase of TLS-limited quality factor and corresponding $T_{1}$ assuming actual parameter values.

\section{Surface Participation Calculation and Optimization method}
\label{Section2}
\subsection{Surface Participation Ratio Calculation}
\begin{figure}[!t]
	\centering
	\includegraphics[width=\columnwidth]{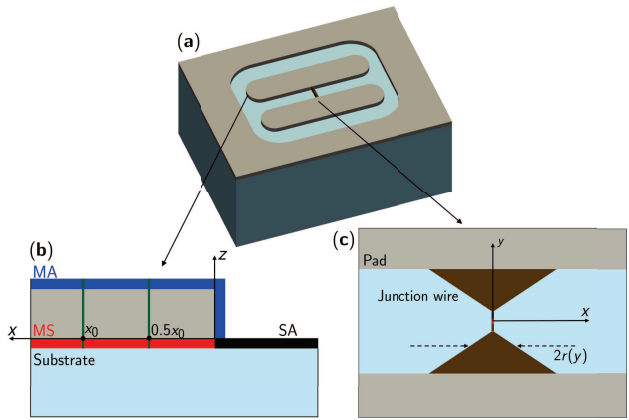}
	\caption{Illustration of TLS layers in planar transmon and junction wire. (a) Schematic of a floating transmon qubit. Two grey-colored capacitor pads are connected with a junction wire represented in brown color. (b) Schematic of TLS layers. The blue, red, and black regions represent the MA, MS, SA interface dielectric layer. Dielectric layers are divided into ``interior'' and ``perimeter'' regions with boundary at a distance $x_0=1\,\mu$m from the metal edge, and the perimeter region is further divided into ``accurate'' region and ``diverging'' region with a boundary at a distance $0.5x_0=0.5\,\mu$m from metal edge. (c) Illustration of junction wire. The red box in the middle represents JJ, brown color represents the junction wire connecting the capacitor pad and JJ, and $r(y)$ denotes the half-width of the junction wire at position $y$.}
	\label{fig_1}
\end{figure}
\indent In low-temperature superconducting circuits, it has been acknowledged that parasitic TLS defects contained in dielectric layers are a noise source of resonators \cite{Gao_semiempirical,Gao_physics} and cause decoherence of superconducting qubit \cite{Martinis_decoherence}. Those TLS-related effects of superconducting circuits can be studied numerically \cite{Wenner_surface,Wang_surface}, and experimentally \cite{Lisenfeld_electric}. In superconducting qubits, a thin dielectric layer is formed at MS (metal-substrate), MA (metal-air), and SA (substrate-air) interface as depicted in \fref{fig_1}-(a),(b), and TLS defects formed in these interface dielectrics are our main interest. Typically, TLS loss of superconducting qubit is studied with the participation ratio model \cite{Wenner_surface,Wang_surface,Melville_comparison,Gambetta_investigating,Calcusine_analysis,Martinis_surface}. In this model, TLS-limited quality factor $Q_{\rm TLS}$ is expressed as,
\begin{equation}
	\label{qtls_eqn}
	Q_{\rm TLS}^{-1}=\sum_{i}p_{i}\,\tan\,\delta_{i},
\end{equation}
\indent where $p_{i}$ and $\tan\delta_{i}$ denotes the geometry-dependent participation ratio and loss tangent of dielectric layer $i$ ($i=$ MS, MA, SA). The participation ratio $p_{i}$ of each dielectric layer can be numerically calculated as \cite{Wenner_surface,Martinis_surface},
\begin{equation}
	\label{pi_eqn}
	p_{i}=\frac{t_{i}\varepsilon_{i}/2}{W}\int_{i}dS\left|E_{i}\right|^{2},
\end{equation}
\indent where $W,t_{i},\varepsilon_{i}$ denotes the total stored energy, thickness, and dielectric constant of $i$'th dielectric layer. In \eref{pi_eqn}, a uniform electric field within a dielectric layer is assumed, and volume integral is replaced with surface integral, which is valid since dielectric interface layers are thin ($\sim$10 nm thickness) compared to the overall transmon geometry feature size ($\sim$100 $\mu$m). However, calculating \eref{pi_eqn} directly from full electromagnetic (EM) simulation is challenging due to the diverging electric field at the metal edge \cite{Wenner_surface,Jackson_classical,Murray_analytical,Sandberg_etch}.\\
\indent For capacitor pads, a common approach to solve this numerical issue is an analytic approach using conformal mapping \cite{Murray_analytical, Martinis_surface}, or using numerical approximations \cite{Wenner_surface, Wang_surface, Wang_towards}, and we adopted the method of \cite{Wang_surface}. As illustrated in \fref{fig_1}-(b), we first divide the capacitor pad into two regions - ``interior'' and ``perimeter''- and the perimeter region is further divided into ``accurate'' region and ``diverging'' region. By assuming that the electric field converges effectively in a pad region distant more than 0.5 $\mu$m from the metal edge, we can numerically calculate the participation ratio of the interior region and accurate region with full EM simulation. In addition, the local field profile nearby the metal edge follows local scaling depending on film thickness \cite{Wang_surface}. Thus, we can calculate the participation of diverging region using scaling factor $F_{i}$, which is defined as the ratio of electric energy stored in accurate, diverging region. \\
\indent Participation ratio of the junction wire was calculated from a separate simulation that only includes the junction wire and nearby pad region, assuming that the participation of junction wire is independent of capacitor pad geometry. This assumption is valid if we maintain the overall dimension of the junction wire much smaller than the capacitor pad feature size. For the junction wire illustrated in \fref{fig_1}-(c), we adopted the flat coax approximation used in \cite{Martinis_surface}, which can effectively approximate the $x$-dependence of the electric field to that of flat coax. Electric energy stored in the upper junction wire can be calculated from flat coax approximation as \cite{Martinis_surface},
\begin{equation}
	\label{pilead_eqn}
	U=t\varepsilon\int_{\rm upper}E(y)^2r(y)\left\lbrace\ln\left(\frac{4r(y)}{t}\right)+5\right\rbrace dy,
\end{equation}
\indent where $E(y)$ and $r(y)$ denotes the electric field and junction wire half-width at the centerline ($x=0$ in \fref{fig_1}-(c)) point with distance $y$ from JJ, $t$ and $\varepsilon$ denotes the thickness and dielectric constant of the dielectric layer, and $5$ in the integrand is a thickness correction factor. Therefore we can calculate the participation ratio from the numerical solution of the electric field along the centerline while assuming perfect conductor boundary condition for superconducting thin films.

\subsection{Optimization method}
\begin{figure}[!b]
	\centering
	\includegraphics[width=0.8\columnwidth]{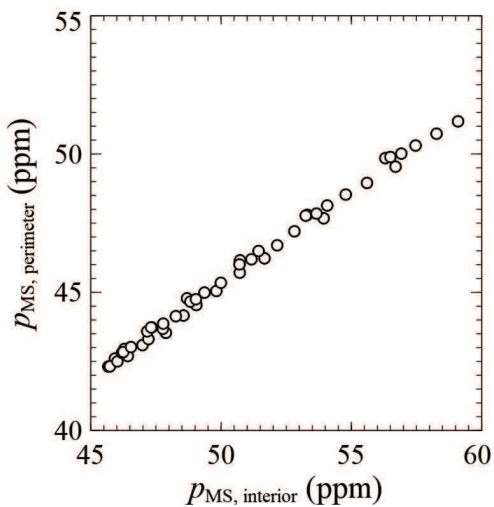}
	\caption{Relation of perimeter, interior participation ratio. $p_{\rm MS}$ of perimeter, interior region of double pad geometry is monitored while varying width, height of capacitor pad.}
	\label{fig_2}
\end{figure}
\begin{figure*}[!t]
	\centering
	\subfloat[]{\includegraphics[width=2.5in]{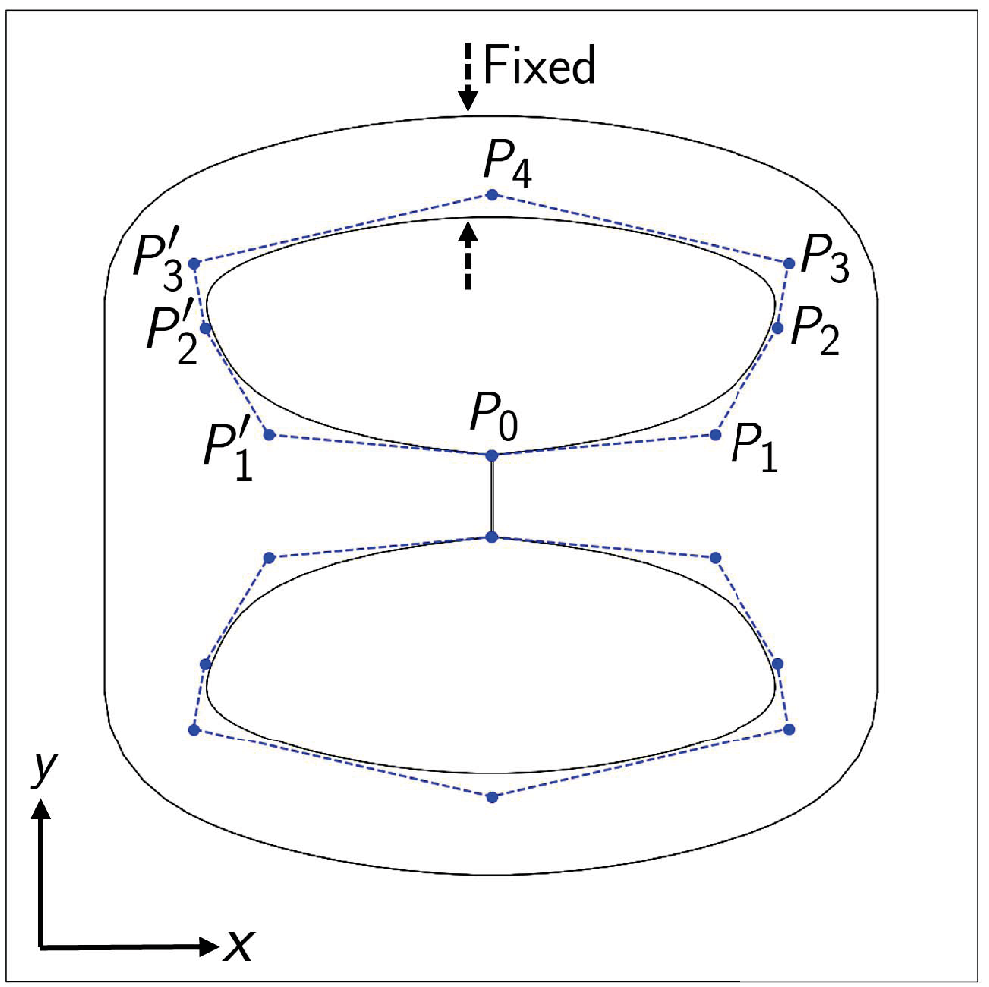}%
		\label{fig_first_case}}
	\hfil
	\subfloat[]{\includegraphics[width=3.5in]{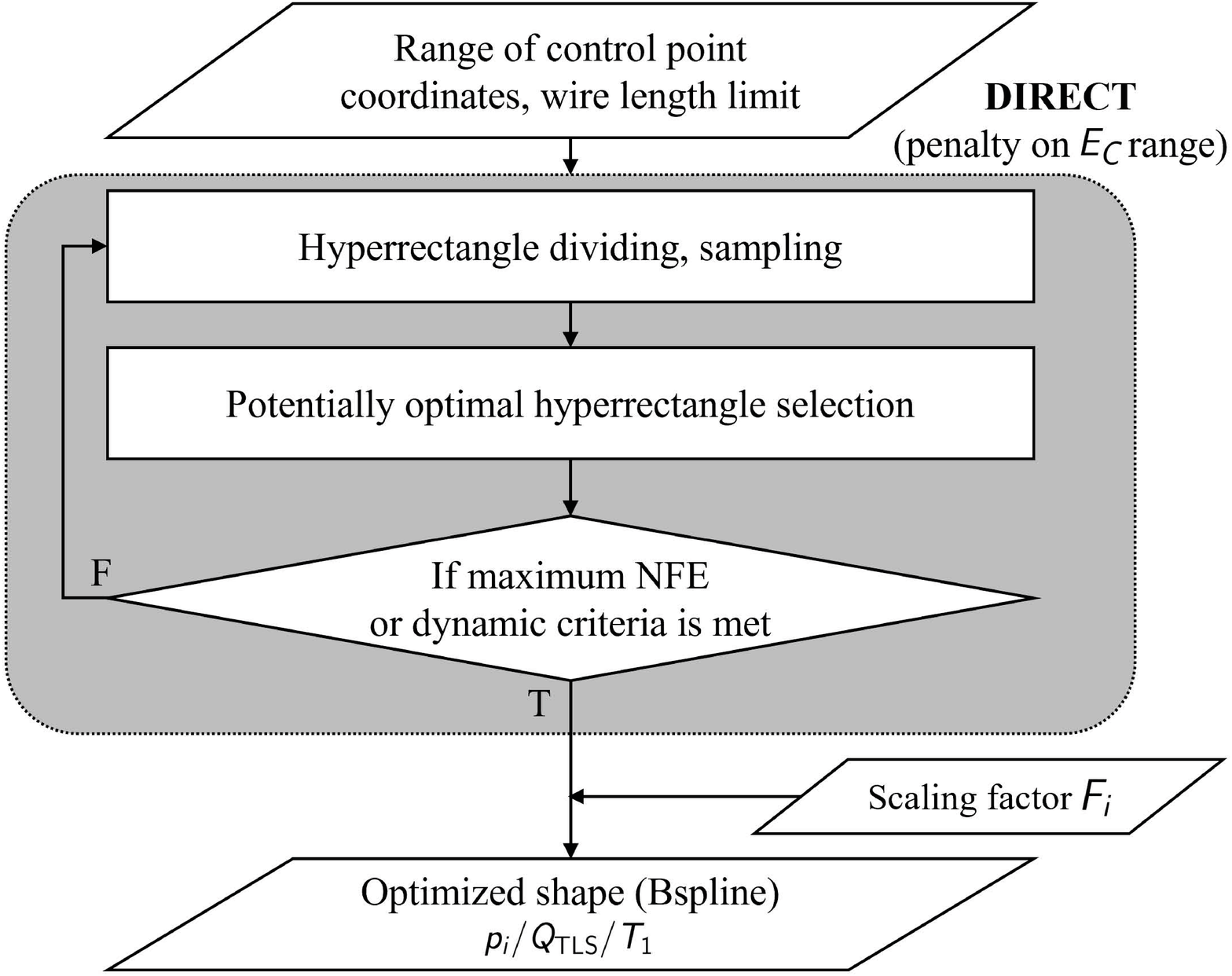}%
		\label{fig_second_case}}
	\caption{Spline-based geometry of capacitor pad and flowchart of the optimization process. (a) For given control point positions ($P_{0}\sim P_{4}$), capacitor pad geometry is constructed using Bspline \cite{Deboor_calculating}. Point $P_{1}',P_{2}'$ and $P_{3}'$ is symmetric to $P_{1},P_{2}$ and $P_{3}$, and the distance between the ground plane and the capacitor pad is fixed. (b) From the given footprint limit and wire length constraint, the range of control point coordinates is determined, and DIRECT optimizer searches design space to find a minimal participation ratio geometry. After each DIRECT iteration, the optimizer checks if the maximum NFE or dynamic criteria is met. As a result, optimized spline-shaped geometry and corresponding participation ratio $p_{i}$ and TLS-limited quality factor $Q_{\rm TLS}$ are obtained.}
	\label{fig_3}
\end{figure*}

\indent Roughly, shape or topology optimization techniques can be categorized into gradient-based and derivative-free optimization methods \cite{Koziel_computational}. For a gradient-based approach, sensitivity analysis is necessary to express the gradient of the objective function in terms of design variables, and it is commonly done using the adjoint variable method \cite{Park_design_book,Koziel_fast}.\\
\indent However, due to numerical reasons, we adopted DIRECT algorithm \cite{Jones_Lipschi,Jones_direct,Gablonsky_locally,Tavassoli_modification} instead, which is a global optimization method capable of finding the optimum region in design space within a few iterations compared to other global optimization methods \cite{Jones_direct,Tavassoli_modification}. During each DIRECT iteration with $N$ total design variables, $N$-dimensional design space is divided into hyperrectangles, and then objective function is evaluated at the center of each divided hyperrectangle. Then, among those hyperrectangles, a set of ``potentially optimal'' samples \cite{Jones_Lipschi} are selected and refined at the following iterations. A detailed description of the algorithm is provided in detail in \cite{Jones_direct}. DIRECT optimization scheme usually terminates when the maximum number of function evaluations (NFE) is reached \cite{Jones_Lipschi,Jones_direct,Gablonsky_locally} or the reduction of the objective function value becomes marginal \cite{Tavassoli_modification}.\\
\indent In capacitor pad geometry optimization, the design variables are the coordinates of control points described in \fref{fig_3}. Total number of design variables is eight, including the $x$ and $y$ coordinates of points $P_{1}$, $P_{2}$, $P_{3}$, and the $y$ coordinates of $P_{0}$ and $P_{4}$ (the total length of the junction wire is determined by the position of $P_{0}$). The objective function was $p_{\rm MS}$ of the interior region. Typically, the participation ratio of interior region and perimeter region move in same direction for transmon qubits with large feature size, verified by separate parametric sweep of double pad geometry in \fref{fig_2}. Each data point in \fref{fig_2} represents simulated pad participation ratio of double pad geometry generated by parametric sweep of capacitor pad dimension. From \fref{fig_2}, $p_{\rm MS}$ of interior, perimeter region moves in same direction and thus the participation ratio of perimeter region can also be minimized by optimizing the participation ratio of interior region for this particular class of floating transmons. For this reason, only the participation ratio of the interior region is included in our objective function. The total participation ratio including the perimeter region is evaluated afterward. The objective function was evaluated with Ansys HFSS eigenmode solver. We used adaptive mesh techniques in which adaptive pass terminates if the difference of $p_{\rm MS}$ between consecutive passes is less than 0.5\%. The optimization process was terminated when the NFE reached 360 iterations, or the following conditions were satisfied.
\begin{eqnarray}
	\label{dyncrit_eqn_1}
	\left|\frac{f_{\rm current}-\bar{f}}{\bar{f}}\right| &\leq 0.5 \cdot 10^{-2},\\
	\label{dyncrit_eqn_2}
	\left|f_{\rm current}-\bar{f}\right| &\leq 0.2 \cdot 10^{-6}.
\end{eqnarray}

\indent Conditions \eref{dyncrit_eqn_1}, \eref{dyncrit_eqn_2} are named as ``dynamic criteria'' in \fref{fig_3}-(b). In \eref{dyncrit_eqn_1} and \eref{dyncrit_eqn_2}, $f_{\rm current}$ denotes the objective function value of potentially optimal hyperrectangles of current DIRECT iteration, and $\bar{f}$ denotes the average function value of the last three iterations. Similar to \cite{Tavassoli_modification}, this dynamic criteria can effectively detect whether the optimization has entered the optimum region and stop the process before it starts excessive refining of design space. Particular limiting value of \eref{dyncrit_eqn_1}, \eref{dyncrit_eqn_2} was selected from typical range of adaptive pass error of finite element simulation \cite{Kosen_building} and the value used in \cite{Tavassoli_modification}. In addition, to maintain a typical range of anharmonicity $\alpha\left(\approx-E_{\rm C}\right)$, $E_{\rm C} \leq350$ MHz condition was forced by introducing a penalty function. We used quadratic penalty function for this constraint \cite{Koziel_computational} as,
\begin{equation}
	\label{penalty_eqn}
	h(\mathbf{x})=\beta\cdot\max\left(0,E_{\rm C}-E_{\rm C,thres}\right)^{2},
\end{equation}
\indent where $\mathbf{x}$ denotes design variables (coordinates of control points), $\beta$ is the adjustable constant, and $E_{\rm C,thres}$ is 350 MHz. In \eref{penalty_eqn}, $E_{\rm C}$ is approximated using,
\begin{equation}
	\label{Ec_eqn}
	hf_{01}\approx\sqrt{8E_{\rm J}E_{\rm C}}-E_{\rm C},
\end{equation}
\indent for transmon regime $E_{\rm J}/E_{\rm C}\gg1$ \cite{Koch_charge}, and junction energy $E_{\rm J}/h$ is assumed to be 16.35 GHz which can be calculated from $L_{\rm J}=10$ nH assumption with the Ambegaokar-Baritoff relation \cite{Ambegaokar_tunneling,Hertzberg_laser}. Using \eref{Ec_eqn}, $E_{\rm C}$ was calculated from the eigenmode solution and fed into the penalty function \eref{penalty_eqn}.\\
\indent In addition, spline-based geometry with five control points was used to optimize junction wire geometry. One of the control points was fixed to the junction location, and the $y$ coordinates of the control points were fixed. Hence there were a total of four design variables. The objective function $p_{\rm MS}$ was calculated from the electric field along centerline $E(y)$. During optimization, $E(y)$ was calculated using electrostatic simulation by applying a differential voltage between the upper and lower junction wires. Then, DIRECT optimization is again used to find a junction wire shape that minimizes the participation ratio, using wire length obtained from capacitor pad optimization. Detailed settings, including parameter setting and termination criteria, are identical to capacitor pad optimization.

\subsection{Parameter Settings and In-Depth Analysis}
\indent In our optimization, we set $\varepsilon_{i}=10$, $t_{i}=3$ nm for dielectric interface layers, which is a simplified setting commonly used in related studies \cite{Wenner_surface,Wang_surface,Martinis_surface}. For further analysis, we first calculated the TLS-limited quality factor and corresponding $T_{1}$ value using actual material parameters measured in the literature \cite{Melville_comparison}. We compared them with existing planar geometries of double pad capacitor design \cite{Gambetta_investigating,Wang_towards,Place_new}, and concentric transmon \cite{Braum_concentric}. Furthermore, we also investigated the effect of a superconductor's surface impedance to validate our simulation's perfect conductor boundary condition. Taking into account the surface impedance of typical aluminum superconducting film, which can be obtained from the Mattis-Bardeen kernel function \cite{Popel_electromagnetic,Zhou_analytical,Gao_physics}, we confirmed that considering surface impedance resulted in participation ratio and resonant frequency change less than 0.1\% at 20 mK. At the same time, it consumes significantly longer computation time. Thus, we safely applied perfect conductor boundary condition for superconducting films.

\section{Results and Discussion}
\label{Section3}
\begin{figure}[!t]
	\centering
	\includegraphics[width=\columnwidth]{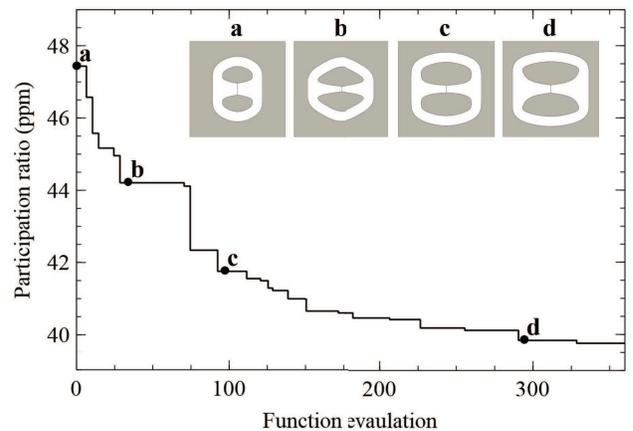}
	\caption{Convergence plot for capacitor pad optimization. Inset (a)-(d) shows how the capacitor pad shape corresponds to each DIRECT iteration's potentially optimal hyperrectangle changes. Y-axis shows the evolution of the capacitor pad interior $p_{\rm MS}$ value of current best sample point.}
	\label{fig_4}
\end{figure}
\begin{figure}[!t]
	\centering
	\includegraphics[width=\columnwidth]{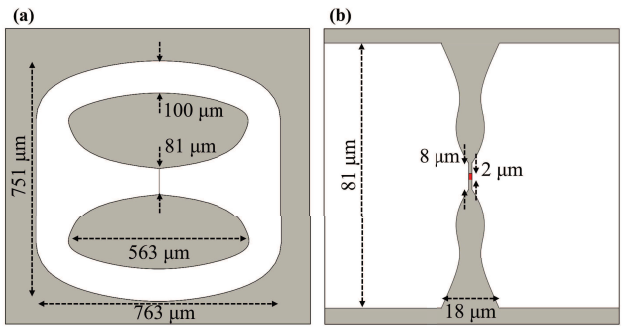}
	\caption{Capacitor pad and junction wire geometry generated by DIRECT optimization. (a) Optimized capacitor pad geometry. The distance between the capacitor pad and the ground plane is fixed to 100 $\mu$m, and resulting overall footprint is 763 $\mu$m $\times$ 751 $\mu$m. (b) Optimized junction wire geometry with wire length of 81 $\mu$m and junction width of 1 $\mu$m. The red box represents JJ.}
	\label{fig_5}
\end{figure}

\begin{figure}[!t]
	\centering
	\includegraphics[width=\columnwidth]{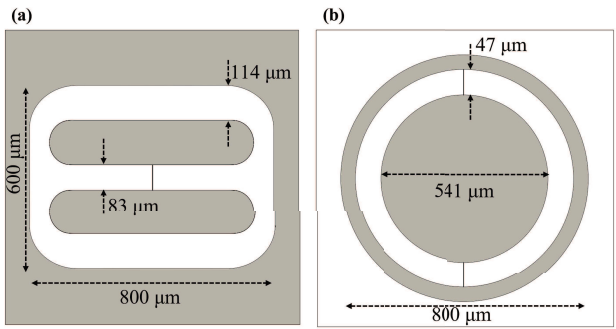}
	\caption{Examples of existing geometries. (a) Double pad capacitor geometry \cite{Wang_towards} with an overall footprint of 800 $\mu$m $\times$ 600 $\mu$m. (b) Concentric transmon \cite{Braum_concentric} with outer diameter 800 $\mu$m.}
	\label{fig_6}
\end{figure}

\indent Optimization of both capacitor pad and junction wire converged within 360 function evaluation with maximum footprint 800 $\mu$m $\times$ 800 $\mu$m. \Fref{fig_4} shows the evolution of the participation ratio in capacitor pad geometry optimization. In \fref{fig_4}, the y-axis shows the capacitor pad interior $p_{\rm MS}$ value of current best sample point. At the early stage of optimization (inset a, b of \fref{fig_4}), $E_{\rm C}/h$ exceeds the set upper bound of 350 MHz. However, after a few iterations, the optimizer approaches the design space region that satisfies $E_{\rm C}$ constraint. After 340 function evaluation, the objective function value converges, and termination conditions are met. DIRECT optimization of junction wire geometry also showed similar convergence and converged in 330 function evaluation. \\
\indent The resulting capacitor pad and junction wire geometries are shown in \fref{fig_5}-(a),(b), respectively. The optimized capacitor geometry in \fref{fig_5}-(a) can be seen as a typical double pad capacitor being smoothly tapered near the junction wire. Similar to the tapered wire suggested in \cite{Martinis_surface}, optimized junction wire geometry \fref{fig_5}-(b) shows tapered wire with a taper slope $S\approx0.4$, with a narrow neck in the middle bringing an additional reduction of participation.

\begin{table}[!t]
	\caption{Key parameters of the optimized capacitor pad, junction wire geometry, and existing structures}
	\centering
	\resizebox{\columnwidth}{!}{\begin{tabular}{@{}*{5}{l}}
			\br
			& & \textbf{Optimized pad} & \textbf{Double pad \cite{Wang_towards}} & \textbf{Concentric \cite{Braum_concentric}}\\
			\mr
			$p_{\rm MS}$ & [ppm] & 81.5 & 97.5 & 106.5\\
			$p_{\rm SA}$ & [ppm] & 70.7 & 82.7 & 88.1\\
			$p_{\rm MA}$ & [ppm] & 6.09 & 7.46 & 8.44\\
			\br
			& & \textbf{Optimized wire} & \textbf{Straight} & \textbf{Linear taper}\\
			\mr
			$p_{\rm MS}$ & [ppm] & 16.8 & 22.8 & 17.8\\
			$p_{\rm MA}$ & [ppm] & 0.17 & 0.23 & 0.18\\
			\br
		\end{tabular}}
	\label{padwireresult_table:1}
	\begin{flushright}
			{\footnotesize
				*$\varepsilon_{i}=10,t_{i}=3$ nm for all dielectric layers
			}
	\end{flushright}
\end{table}

\indent Key parameters of the optimized capacitor pad, junction wire, and some reference geometries are extracted from the EM simulation and summarized in \tref{padwireresult_table:1}. \Fref{fig_6} shows the geometry and dimension of the existing structures. In \tref{padwireresult_table:1}, total MS, MA participation ratio was calculated using scaling factor $F_{i}$, which was determined by 2D local electrostatic simulation of the thin metal film with a thickness of 0.1 $\mu$m. For double pad capacitor geometry, we used the one in \cite{Wang_towards}, and concentric transmon geometry has an outer diameter of 800 $\mu$m to match the overall footprint of our design. \Tref{padwireresult_table:1} shows that the total participation ratio of MS, MA, SA was reduced by approximately $15\sim18$\% compared to double pad capacitor geometry. The lower part of \tref{padwireresult_table:1} shows extracted participation ratio of resulting junction wire geometry, straight wire, and linear taper with taper slope $S=0.4$. Linear taper and optimized geometry display a participation ratio of approximately $22\sim26$\% smaller than straight wire, and the participation ratio of optimized wire is slightly smaller than linear taper.\\
\indent Based on the parameters in \tref{padwireresult_table:1}, TLS-limited quality factor $Q_{\rm TLS}$ and corresponding TLS-limited relaxation time $T_{1}\left(= Q_{\rm TLS}/\omega_{01}\right)$ were calculated using realistic parameters. To demonstrate, we assumed niobium on a silicon substrate device \cite{Niepce_geometric}. We used material parameters reported in \cite{Niepce_geometric} ($\tan\,\delta_{\rm MS}=1.3\times 10^{-3}$, $\tan\,\delta_{\rm SA}=2.1\times 10^{-3}$, $\tan\,\delta_{\rm MA}=4.7\times 10^{-2}$) while assuming dielectric constant $\varepsilon_{\rm MS}=11.7$, $\varepsilon_{\rm sub}=11.7$, $\varepsilon_{\rm SA}=4.2$, $\varepsilon_{\rm MA}=33$ and dielectric layer thickness $t_{\rm MS}=2$ nm, $t_{\rm SA}=5$ nm, $t_{\rm MA}=5$ nm. TLS-limited quality factor and $T_{1}$ of transmon qubit with double pad capacitor shape and straight wire were calculated as 2.24$\times10^{6}$ and 71.1 $\mu$s assuming qubit frequency $f_q=5$ GHz. On the other hand, $Q_{\rm TLS}$ and $T_{1}$ of transmon qubit with optimized capacitor pad and junction wire geometry were 2.72$\times10^{6}$ and 86.5 $\mu$s. Hence, both TLS-limited quality factor and relaxation time were improved by 21.6\% through shape optimization.

\section{Conclusion}
\indent We present a shape optimization-based approach to improve the $T_{1}$ of superconducting transmon qubits. A common approach for the reduction of surface dielectric loss is changing material or fabrication method, but we adopted geometry optimization techniques that were commonly used in other electromagnetic systems. The results indicate that surface dielectric loss can be reduced by optimizing the geometry of transmon qubit. Further research of geometry optimization-based approach of surface dielectric loss reduction is required to design a qubit structure with smaller footprints yet maintaining sufficient $T_{1}$. Moreover, combining geometry optimization-based approach with recent tantalum transmon qubits \cite{Place_new,Wang_towards} will further improve the state-of-the-arts transmon qubits, and studying other loss mechanisms, including quasiparticle loss  \cite{Lutchyn_kinetics} will widen the understanding of the optimized geometry. Our transmon qubit design can also be applied to realizing practical quantum computer since longer $T_{1}$ reduces coherence limited quantum gate error \cite{Abad_universal}.

\ack{This research was supported by 2022 Student-Directed Education Regular Program from Seoul National University and National R\&D Program through the National Research Foundation of Korea(NRF) funded by Ministry of Science and ICT(2022M3I9A1072846) and in part by the Applied Superconductivity Center, Electric Power Research Institute of Seoul National University.}

\section*{Conflict of interest}
The authors declare no conflict of interest.

\section*{References}
\bibliographystyle{iopart-num}
\bibliography{reflist}

\providecommand{\newblock}{}
\begin{thebibliography}{10}
\expandafter\ifx\csname url\endcsname\relax
  \def\url#1{{\tt #1}}\fi
\expandafter\ifx\csname urlprefix\endcsname\relax\def\urlprefix{URL }\fi
\providecommand{\eprint}[2][]{\url{#2}}

\bibitem{Devoret_superconducting}
Devoret M~H and Schoelkopf R~J 2013 {\em Science\/} {\bf 339} 1169--1174

\bibitem{Arute_quantum}
Arute F, Arya K, Babbush R, Bacon D, Bardin J~C, Barends R, Biswas R, Boixo S,
  Brandao F~G and Buell D~A 2019 {\em Nature\/} {\bf 574} 505--510

\bibitem{Koch_charge}
Koch J, Terri M~Y, Gambetta J, Houck A~A, Schuster D~I, Majer J, Blais A,
  Devoret M~H, Girvin S~M and Schoelkopf R~J 2007 {\em Phys. Rev. A\/} {\bf 76}
  042319

\bibitem{Krantz_quantum}
Krantz P, Kjaergaard M, Yan F, Orlando T~P, Gustavsson S and Oliver W~D 2019
  {\em Appl. Phys. Rev.\/} {\bf 6} 021318

\bibitem{Anderson_anomalous}
Anderson P~W, Halperin B~I and Varma C~M 1972 {\em Philos. Mag.\/} {\bf 25}
  1--9

\bibitem{Martinis_decoherence}
Martinis J~M, Cooper K~B, McDermott R, Steffen M, Ansmann M, Osborn K, Cicak K,
  Oh S, Pappas D~P and Simmonds R~W 2005 {\em Phys. Rev. Lett.\/} {\bf 95}
  210503

\bibitem{Gao_semiempirical}
Gao J, Daal M, Martinis J~M, Vayonakis A, Zmuidzinas J, Sadoulet B, Mazin B~A,
  Day P~K and Leduc H~G 2008 {\em Appl. Phys. Lett.\/} {\bf 92} 212504

\bibitem{Gao_experimental}
Gao J, Daal M, Vayonakis A, Kumar S, Zmuidzinas J, Sadoulet B, Mazin B~A, Day
  P~K and Leduc H~G 2008 {\em Appl. Phys. Lett.\/} {\bf 92} 152505

\bibitem{Muller_towards}
M{\"u}ller C, Cole J~H and Lisenfeld J 2019 {\em Rep. Prog. Phys.\/} {\bf 82}
  124501

\bibitem{Geerlings_improving}
Geerlings K, Shankar S, Edwards E, Frunzio L, Schoelkopf R and Devoret M 2012
  {\em Appl. Phys. Lett.\/} {\bf 100} 192601

\bibitem{Lisenfeld_electric}
Lisenfeld J, Bilmes A, Megrant A, Barends R, Kelly J, Klimov P, Weiss G,
  Martinis J~M and Ustinov A~V 2019 {\em Npj Quantum Inf.\/} {\bf 5} 1--6

\bibitem{Wenner_surface}
Wenner J, Barends R, Bialczak R, Chen Y, Kelly J, Lucero E, Mariantoni M,
  Megrant A, O’Malley P and Sank D 2011 {\em Appl. Phys. Lett.\/} {\bf 99}
  113513

\bibitem{Wang_surface}
Wang C, Axline C, Gao Y~Y, Brecht T, Chu Y, Frunzio L, Devoret M and Schoelkopf
  R~J 2015 {\em Appl. Phys. Lett.\/} {\bf 107} 162601

\bibitem{Melville_comparison}
Melville A, Calusine G, Woods W, Serniak K, Golden E, Niedzielski B~M, Kim D~K,
  Sevi A, Yoder J~L and Dauler E~A 2020 {\em Appl. Phys. Lett.\/} {\bf 117}
  124004

\bibitem{Gambetta_investigating}
Gambetta J~M, Murray C~E, Fung Y~K~K, McClure D~T, Dial O, Shanks W, Sleight
  J~W and Steffen M 2016 {\em IEEE Trans. Appl. Supercond.\/} {\bf 27} 1--5

\bibitem{Calcusine_analysis}
Calusine G, Melville A, Woods W, Das R, Stull C, Bolkhovsky V, Braje D, Hover
  D, Kim D~K and Miloshi X 2018 {\em Appl. Phys. Lett.\/} {\bf 112} 062601

\bibitem{Martinis_surface}
Martinis J~M 2022 {\em Npj Quantum Inf.\/} {\bf 8} 1--12

\bibitem{Wang_towards}
Wang C, Li X, Xu H, Li Z, Wang J, Yang Z, Mi Z, Liang X, Su T and Yang C 2022
  {\em Npj Quantum Inf.\/} {\bf 8} 1--6

\bibitem{Place_new}
Place A~P, Rodgers L~V, Mundada P, Smitham B~M, Fitzpatrick M, Leng Z,
  Premkumar A, Bryon J, Vrajitoarea A and Sussman S 2021 {\em Nat. Commun.\/}
  {\bf 12} 1--6

\bibitem{Deng_titanium}
Deng H, Song Z, Gao R, Xia T, Bao F, Jiang X, Ku H~S, Li Z, Ma X and Qin J 2022
  {\em arXiv preprint arXiv:2205.03528\/}

\bibitem{Dial_bulk}
Dial O, McClure D~T, Poletto S, Keefe G, Rothwell M~B, Gambetta J~M, Abraham
  D~W, Chow J~M and Steffen M 2016 {\em Supercond. Sci. Technol.\/} {\bf 29}
  044001

\bibitem{Park_design}
Park S~H, Bang J, An S and Hahn S 2021 {\em IEEE Trans. Appl. Supercond.\/}
  {\bf 31} 1--5

\bibitem{Linsenfeld_decoherence}
Lisenfeld J, Bilmes A, Matityahu S, Zanker S, Marthaler M, Schechter M, Schön
  G, Shnirman A, Weiss G and Ustinov A~V 2016 {\em Sci. Rep.\/} {\bf 6} 1--8

\bibitem{Park_design_book}
Park I~H 2019 {\em Design Sensitivity Analysis and Optimization of
  Electromagnetic Systems\/} (Springer)

\bibitem{Qian_topology}
Qian X 2013 {\em Comput. Methods Appl. Mech. Eng.\/} {\bf 265} 15--35

\bibitem{Allaire_structural}
Allaire G, Jouve F and Toader A~M 2004 {\em J. Comput. Phys.\/} {\bf 194}
  363--393

\bibitem{Deng_self}
Deng C, Wang Y, Qin C, Fu Y and Lu W 2022 {\em Nat. Commun.\/} {\bf 13} 1--14

\bibitem{Koziel_fast}
Koziel S and Bekasiewicz A 2015 {\em IEEE Antennas Wireless Propag. Lett.\/}
  {\bf 14} 1681--1684

\bibitem{John_antenna}
John M and Ammann M~J 2009 {\em IEEE Trans. Antennas Propag.\/} {\bf 57}
  260--263

\bibitem{Zhou_level}
Zhou S, Li W and Li Q 2010 {\em J. Comput. Phys.\/} {\bf 229} 6915--6930

\bibitem{Lizzi_optimization}
Lizzi L, Viani F, Azaro R and Massa A 2007 {\em IEEE Antennas Wireless Propag.
  Lett.\/} {\bf 6} 182--185

\bibitem{Lizzi_PSO}
Lizzi L, Viani F, Azaro R and Massa A 2008 {\em IEEE Trans. Antennas Propag.\/}
  {\bf 56} 2613--2621

\bibitem{Hassan_topology}
Hassan E, Wadbro E and Berggren M 2014 {\em IEEE Trans. Antennas Propag.\/}
  {\bf 62} 2488--2500

\bibitem{Toivanen_gradient}
Toivanen J~I, M{\"a}kinen R~A, Rahola J, J{\"a}rvenp{\"a}{\"a} S and
  Yl{\"a}-Oijala P 2010 {\em IET Microw. Antennas Propag.\/} {\bf 4} 1406--1414

\bibitem{Koulouridis_novel}
Koulouridis S and Volakis J~L 2008 {\em IEEE Antennas Wireless Propag. Lett.\/}
  {\bf 8} 34--36

\bibitem{Lee_continuum}
Lee K~H, Choi C~Y and Park I~H 2017 {\em IEEE Trans. Magn.\/} {\bf 54} 1--4

\bibitem{Georgieva_feasible}
Georgieva N~K, Glavic S, Bakr M~H and Bandler J~W 2002 {\em IEEE Trans. Microw.
  Theory Techn.\/} {\bf 50} 2751--2758

\bibitem{Weile_genetic}
Weile D~S and Michielssen E 1997 {\em IEEE Trans. Antennas Propag.\/} {\bf 45}
  343--353

\bibitem{Misilmani_review}
El~Misilmani H~M, Naous T and Al~Khatib S~K 2020 {\em Int. J. RF Microw.
  Comput.-Aided Eng.\/} {\bf 30} e22356

\bibitem{Gao_physics}
Gao J 2008 {\em The physics of superconducting microwave resonators\/} Thesis
  California Institute of Technology

\bibitem{Jackson_classical}
Jackson J~D 1999 {\em Classical electrodynamics\/} 3rd ed (Wiley)

\bibitem{Murray_analytical}
Murray C~E, Gambetta J~M, McClure D~T and Steffen M 2018 {\em IEEE Trans.
  Microw. Theory Techn.\/} {\bf 66} 3724--3733

\bibitem{Sandberg_etch}
Sandberg M, Vissers M~R, Kline J~S, Weides M, Gao J, Wisbey D~S and Pappas D~P
  2012 {\em Appl. Phys. Lett.\/} {\bf 100} 262605

\bibitem{Deboor_calculating}
De~Boor C 1972 {\em Journal of approximation theory\/} {\bf 6} 50--62

\bibitem{Koziel_computational}
Koziel S and Yang X~S 2011 {\em Computational optimization, methods and
  algorithms\/} vol 356 (Springer)

\bibitem{Jones_Lipschi}
Jones D~R, Perttunen C~D and Stuckman B~E 1993 {\em J. Optim. Theory Appl.\/}
  {\bf 79} 157--181

\bibitem{Jones_direct}
Jones D~R and Martins J~R 2021 {\em J. Glob. Optim.\/} {\bf 79} 521--566

\bibitem{Gablonsky_locally}
Gablonsky J~M and Kelley C~T 2001 {\em J. Glob. Optim.\/} {\bf 21} 27--37

\bibitem{Tavassoli_modification}
Tavassoli A, Haji~Hajikolaei K, Sadeqi S, Wang G~G and Kjeang E 2014 {\em
  Optim. Eng.\/} {\bf 46} 810--823

\bibitem{Kosen_building}
Kosen S, Li H~X, Rommel M, Shiri D, Warren C, Grönberg L, Salonen J, Abad T,
  Biznárová J and Caputo M 2022 {\em Quantum Sci. Technol.\/} {\bf 7} 035018

\bibitem{Ambegaokar_tunneling}
Ambegaokar V and Baratoff A 1963 {\em Phys. Rev. Lett.\/} {\bf 10} 486

\bibitem{Hertzberg_laser}
Hertzberg J~B, Zhang E~J, Rosenblatt S, Magesan E, Smolin J~A, Yau J~B, Adiga
  V~P, Sandberg M, Brink M and Chow J~M 2021 {\em Npj Quantum Inf.\/} {\bf 7}
  1--8

\bibitem{Braum_concentric}
Braumüller J, Sandberg M, Vissers M~R, Schneider A, Schlör S, Grünhaupt L,
  Rotzinger H, Marthaler M, Lukashenko A and Dieter A 2016 {\em Appl. Phys.
  Lett.\/} {\bf 108} 032601

\bibitem{Popel_electromagnetic}
P{\"o}pel R 1989 {\em Electromagnetic properties of superconductors\/}
  (Springer) pp 44--78

\bibitem{Zhou_analytical}
Zhou S~p, Jabbar A, Bao J~S, Wu K~q and Jin B~j 1992 {\em J. Appl. Phys.\/}
  {\bf 71} 2789--2794

\bibitem{Niepce_geometric}
Niepce D, Burnett J~J, Latorre M~G and Bylander J 2020 {\em Supercond. Sci.
  Technol.\/} {\bf 33} 025013

\bibitem{Lutchyn_kinetics}
Lutchyn R, Glazman L and Larkin A 2006 {\em Phys. Rev. B\/} {\bf 74} 064515

\bibitem{Abad_universal}
Abad T, Fern{\'a}ndez-Pendás J, Kockum A~F and Johansson G 2022 {\em Phys.
  Rev. Lett.\/} {\bf 129} 150504

\end{thebibliography}

\end{document}